\documentstyle[preprint,aps]{revtex}
\begin{document}
\draft
 
\title{Randomly connected cellular automata: A search for critical
connectivities}

\author{Normand Mousseau}
 
\address{Theoretical Physics, University of Oxford \\
1 Keble Road, Oxford, U.K. OX1 3NP }
\date{\today}
\maketitle

\begin{abstract}
I study the Chat{\'e}--Manneville cellular automata rules on randomly connected
lattices. The periodic and quasi--periodic macroscopic behaviours
associated with these rules on finite--dimensional lattices persist on
an infinite--dimensional lattice with finite connectivity and symmetric bonds. 
The lower critical connectivity for these models is at $C=4$ and the
mean--field connectivity, if finite, is not smaller than $C=100$.
Autocorrelations are found to decay as a power--law with a connectivity
independent exponent $\sim -2.5$.  A new intermittent
chaotic phase is also discussed. 
\end{abstract}

\pacs{PACS: 05.50, 05.45, 03.20}
 
\section*{  }

There has been some interest, recently, regarding the existence of spatially
homogeneous systems displaying periodic or quasi--periodic temporal
oscillations \cite{chate91,hemmin92,grinstein93,herz95,bottani95}.  This
interest is motivated both by the desire to develop a better
understanding of complex dynamical systems and by the need to find simple
models representing with some faithfulness dynamical oscillations found in
biological and neurophysiological systems \cite{herz95,mouches}. In spite
of the fair amount of work already accomplished, the fundamental mechanism
at play in simple models like the Chat{\'e}--Manneville \cite{chate91} or the 
Hemmingsson \cite{hemmin92} cellular automata (CA) remains elusive and 
satisfactory 
analytical solutions are still missing. I present here a study of the
Chat{\'e}--Manneville rules on randomly--connected lattices. 
These lattices present many advantages.  By placing CA on them, we move from a
finite--dimensional problem to an infinite one with {\it finite}
connectivity. The concept of space, therefore, becomes irrelevant. This
removes dimensionality along with the questions of lattice--related
effects, choice of neighbours, etc.,  from the problem, leaving a
single parameter: the connectivity.  This simplification puts a
restriction on the necessary and sufficient conditions to obtain
global (quasi--)periodicity. Randomly connected lattices are also closer
to  many globally periodic biological systems, like neurons or fireflies
\cite{herz95,bottani95}, for example, than the ordered ones.

In this letter, I study the effect of connectivity on the global behaviour
of the Chat{\'e}--Manneville type CA.  In particular, I search for lower 
and higher critical connectivities.  A lower
critical coordination, at which the first cyclic phase  appear,
 is found at $C=4$, a smaller value than
what could have been expected from the work on finite--dimensional
lattices \cite{chate92}.  As regard the upper critical limit, 
where a mean--field solution would apply,  
even with a connectivity as high as $100$ quasi--periodic phases can still
be found. However, the number of CA rules leading to  non--trivial phases
peaks at relatively low connectivity, $C=13$,
and decreases rapidly as one moves towards higher number of neighbours. 

The rules used here were proposed by Chat{\'e} and Manneville a few years ago
\cite{chate91}. Since their introduction, these CA have been extensively
studied on finite--dimensional lattices but they remain imperfectly 
understood (see, for example, \cite{chate92,gallas92,mousseau94}). 
I refer the reader to these papers for more details and
will give here only a brief description of the rules. These CA are
two-state objects incremented with a parallel dynamics following a
totalistic rule given by
\begin{eqnarray}
s_i(t+1) = \left\{ \begin{array}{ll}
    1 & \mbox{if $ S_{min} \leq s_i(t) + \sum_{ j \in {\cal N}_i} s_j(t)
\leq S_{max} $ }\\
    0 & \mbox{otherwise} \end{array} \right.
\end{eqnarray}
where ${\cal N}_i$ is the neighbourhood of site $i$.  In the rest of this
paper, I shall use ${\cal R}^C_{S_{min}-S_{max}}$ to denote the rule used,
where $C$ is the connectivity of the randomly connected lattice.  The
macroscopic quantity focused on is mainly the concentration of sites in
state $1$, $c(t) = 1/N \sum_i s_i(t)$.

The construction of a randomly--connected lattice can be pursued following
different procedures.  If oriented bonds are distributed randomly 
between the various sites such that if
$i$ interacts with  $j$, $j$ does not necessarily interact with $i$ (in
the thermodynamical limit, with finite connectivity, the probability of
$j$ being also a neighbour of $i$ is zero), the macroscopic quantity $c(t)$
is well described by a second order mean--field solution which
takes into account correlations between pairs of nearest--neighbour
sites. This result has already been discussed by Chat{\'e} and Manneville
\cite{chate92}.

Requiring that the bonds be symmetric changes completely this picture. 
The resulting macroscopic behaviour is strongly non--mean--field like and
resemble closely the type of behaviour found on the ordered lattices.  In
the thermodynamical limit, a randomly connected lattice with a finite
connectivity is strictly equivalent to a Cayley tree, i.e. that   
there exists
only a single path between any two points on the network. Because of finite
size, this is not true for computer simulations. Keeping 
open boundaries at the edges is not feasible because on a Cayley tree the number
of sites at the boundary is roughly equal to the number of sites in the bulk. 
Randomly connected lattices are 
constructed instead by drawing a list of neighbours at random with only two
restrictions: a fixed coordination and single bond between any two sites.
Doing so, loops of length three and up are implicitly built in the
network.  However, the lack of symmetry (i.e. loops of almost any length
are present) should reduce significantly the finite--size effects on the
general properties of the CA.  To ascertain the validity of this approach,
I have performed a series a simulations on networks of different sizes following
rule ${\cal R}^4_{2-4}$, from an average loop length of $5$ (256 sites) to
$12$ (more that 4 million sites). Results show that the temporal behaviour
is not affected by the system size except for the usual macroscopic noise which
decreases with the number of sites in the model \cite{chate92}.  
Although in this work networks with a  high connectivity can have 
an average loop length as low as 4, the preceding results indicate 
that the behaviour obtained should correspond, at least qualitatively, 
to what would be found in the thermodynamical limit.
 
We concentrate first on low and high connectivities.
For connectivities from $C=2$ to $C=36$, all the $C(C+1)/2$ rules were
examined starting from a random initial configuration with an equal number of
sites in state $0$ and $1$. For larger coordination, only the regions in
the $S_{min}-S_{max}$ plane where non--trivial rules were found at $C=36$
were examined, i.e.  around $S_{min}=S_{max} = C/2$.  At very low
coordination, $C = 2$ and 3, CA do not display any periodic or
quasi-periodic cycles.  However, for certain rules, the models possess
noisy fixed points ($P1$). The first cyclic phase
appears at $C=4$, where a stable quasi--periodic cycle (rule 
${\cal R}^4_{2-4}$)  with a period close to 3 ({\it QP}3) is found (see Fig.
\ref{fig:4c}). This finding supports claims by Grinstein {\it et
al.}\cite{grinstein93} that on finite-dimensional lattices, it is
not the low connectivity generally associated with low dimensions which
prevents quasi--periodic cycles to stabilise but really the topology of
the space.  It shows, also, that finite dimensionality or 
a symmetric lattice are not necessary to obtain periodicity or
quasi--periodicity. We note that the transient period
here is extremely short: with a 4 million site network, starting with
$c(t_0=0) = 0.50$, the CA falls on its cyclic attractor in one time step
only. This very short transient is also found for many other rules and
connectivities.

The question of the existence of an upper critical dimension for ordered
lattices remains unanswered in part because of the lack of a proper
theory for these CA and in part because it is numerically difficult, for
storage reasons, to go to dimensions higher than 8 or 10. On a randomly
connected lattice, it is much easier to achieve very large connectivity
because there is no minimum size for the unit cell. 

The passage to a mean--field like time evolution
appears slowly as one increases the coordination number. Although
certain rules still display non--trivial behaviour at $C=22$, others, 
${\cal R}^{22}_{1-2}$ for
example (Fig. \ref{fig:mf}), follow closely, albeit not exactly, the mean--field
solution,  
\begin{equation}
c(t+1)  = \sum_{r = S_{min}}^{S_{max}} \frac{(C+1)!}{r! (C+1-r)!} c(t)^r
[1-c(t)]^{C+1-r}.
\label{eq:mf}
\end{equation}
Because of finite size effects, there exists a threshold for $c(t)$ in the
simulation under which the CA will always go to zero. In order to
compare the mean--field solution with the simulations, it is useful to
introduce a similar cut--off in equation \ref{eq:mf}.
Doing so, it turns out that, for large connectivities, the mean--field 
rules which do not cross this threshold and, therefore, retain a
non--zero $c((t)$ are the only ones leading to non--trivial cycles in
the simulation. 

However, even reaching connectivities as large as $C=100$, these
non--vanishing rules, with
$S_{min}$ just under and $S_{max}$ just over $C/2$, still display
non--mean field behaviour ($P1$ and {\it QP}3), indicating that if there is a
mean-field connectivity, it can only be found at very high coordination.
Since it is difficult, for technical reason, to simulate a CA with  a 
connectivity much larger
than $C=100$, we have to extract information by following the change
in the number and behaviour of non--trivial rules as one varies the
coordination number.  With increasing coordination, the size of the $QP3$
cycle shrinks from a width of about $0.22$ with $C=4$ to about $0.03$ with
$C=100$.  Similarly, the center of the cycle moves from $c(t) \sim 0.70$
to $0.49$.  As shown in Figure \ref{fig:conn}, the number of non--trivial
rules, $n(C)$,  for a given connectivity ---i.e. rules that are not
locally periodic nor in the absorbing state zero--- reaches a maximum for
$C=13$.  It increases almost linearly from $C=2$ and decreases, for large
connectivities, roughly as a power--law $n(C) \sim (C-12)^{-0.75}$. For
$C=100$, only two rules out of $5050$ are found to give non--trivial
behaviour. If we use this power--law fit as an indicator, the upper
critical connectivity should be in the 100s-range at its lower value.
The existence of such a critical value would appear, again from this
power--law decrease, to be an artefact of the discrete nature of the
rules. If $S_{min}$ and $S_{max}$ were continuous, the non--trivial area,
in this two--dimensional space, would simply decrease continuously to
reach zero only for infinite connectivity.

The type of behaviour found in the $S_{min}$--$S_{max}$ diagram for a 
connectivity between, say, 8 and 14 neighbours is very close to what was 
found by Chat{\'e} and Manneville in 4 to 6 dimensions \cite{chate92}. The
zones identified by these two authors as concentrating $P1$, $P2$ and more
complex time behaviour are in extremely good agreement with what is found
on these infinite dimensional lattices. However, we see  that 
the $S_{min}$--$S_{max}$ space does not become richer as
the connectivity increases beyond $C=13$.  From \cite{chate95}, we
know that spatial correlation falls as $ r_{ij}^{2-d}$, with $r_{ij}$
the distance between two sites. For large $d$,
the decay will be so fast that cycles should be dominated by 
connectivity and not dimensionality.  Therefore, the discussion of
this paragraph is expected to hold also for finite dimensional
lattices: contrary to the  predictions of Chat\'e and
Manneville \cite{chate92}, the number of non--trivial
rules should become very small in high dimensions. 
 
At low coordination, the randomly--connected CA presents a
type of phase which is not found on the ordered lattice:
intermittent--chaotic phases displaying a very complex Poincar{\'e} map.
Rule ${\cal R}^{10}_{2-6}$, for example, intermittently goes from a
periodic two to a wide-band chaotic behaviour (Fig. \ref{fig:chaos}(a)).
 Moreover, as can be seen from Fig. \ref{fig:chaos}(b), this cycle not 
belong to mean--field solutions. 

As regard the stability of the phases discussed here, 
introduction of external noise and internal frustration show that
they are as stable as the CA on an ordered lattice. External noise was
introduced following the protocol described in \cite{mousseau95} and 
cycles persist with finite amount of random spin inversion after each
timestep. The effect of frustration will be discussed elsewhere but 
cycles remain stable under finite amount of it. 

Finally, a word about the decay in the site auto--correlation function,
\begin{equation}
S(t,t_0) = \frac{1}{N} \sum_i s_i(t)s_i(t_0) - c(t)c(t+1).
\end{equation}
It has been recently proposed that for finite--dimensional CA, 
rules producing
quasi-periodic cycles show a power-law decay  going as
$(t-t_0)^{-(d-2)/2}$. For the randomly--connected lattice, 
autocorrelations also decay as a power law. However,  the exponent,
 $\simeq -2.5$, seems to be independent of the connectivity  
as well as the type of non--trivial cyclic behaviour (Fig. \ref{fig:corr}).
The meaning of this exponent still remains to be explained. 

In this letter, I have established a lower critical connectivity for
the Chat{\'e}--Manneville--type CA on a randomly--connected lattice. 
Results seem to indicate also the presence of a higher critical connectivity,
due to the discreteness of the problem, which should be in the region of
a few hundred neighbours. 
From work on finite--dimensional lattices, it was conjectured that the
complexity of the cycles would increase with the dimensionality of the
space to break down only when reaching $d=\infty$ \cite{chate92}. This
ever--increasing complexity does not seem to happen. There is a
peak in the number of non--trivial phases at a finite connectivity
$C=13$ followed by a rapid decrease in this number as connectivity is 
increased.  Because for high dimensional lattices the spatial 
correlation decay is expected to be fast, results obtained here should be
qualitatively applicable to those in this limit. 
 
These results improve on the general understanding about what are the 
fundamental ingredients
needed to obtain non--trivial periodic or quasi--periodic behaviour.  In
particular, (1) symmetric bonds are necessary to obtain non--mean field
behaviour. However, no other loops are needed. (2) Particular topology or
symmetries for the network are not needed.  (3) The lower connectivity is
$4$ while the higher critical one should be very large (at least, a few
hundred neighbours), if is exists at all. These results could also be
applied to biological problems where the effective dimensionality is high.

\section*{Acknowledgements}
 
I would like to acknowledge helpful discussions with G.T. Barkema and D.
Sherrington.  I would also like to thank the Natural Sciences and
Engineering Research Council of Canada for a post-doctoral fellowship.

\begin{figure}
\caption{Magnetisation map for an automata following ${\cal R}^4_{2-4}$.
One thousand timesteps on a system with 4 194 304 sites. The transient
here is one timestep.}
\label{fig:4c}
\end{figure}

\begin{figure}
\caption{Number of rules leading to non--trivial time behaviour as a
function of connectiviy. A non--trivial rules is one which does not lead
the system to zero or local periodic states. The lines are guide to the
eyes and are described in the text.}
\label{fig:conn}
\end{figure}

\begin{figure}
\caption{ (a) Time evolution of the macroscopic $c(t)$ for intermittent
chaotic--periodic rule 
${\cal R}^{10}_{2-6}$;
$500 000$ sites, $20000$ timesteps.  (b) Poincar{\'e} map 
for the same rule. The line is the mean-field solution.}
\label{fig:chaos}
\end{figure}

\begin{figure}
\caption{For large connectivity, the magnetisation map is close
to the mean--field solution. Here, ${\cal R}^{22}_{2-12}$ on a 500
000--sites lattice. The line is the mean--field solution.}
\label{fig:mf}
\end{figure}

\begin{figure}
\caption{Auto-correlation as a function of time for ${\cal R}^4_{2-4}$
(short dashes), ${\cal R}^{10}_{2-4}$ (solid line) and ${\cal
R}^{22}_{9-12}$ (long dashes), respectively a {\it QP}3, $P$2$\times${\it
QP}3 and {\it QP}3 cycles. The straight line is a guide to the eye,
with a slope of $-2.5$.}
\label{fig:corr}
\end{figure}

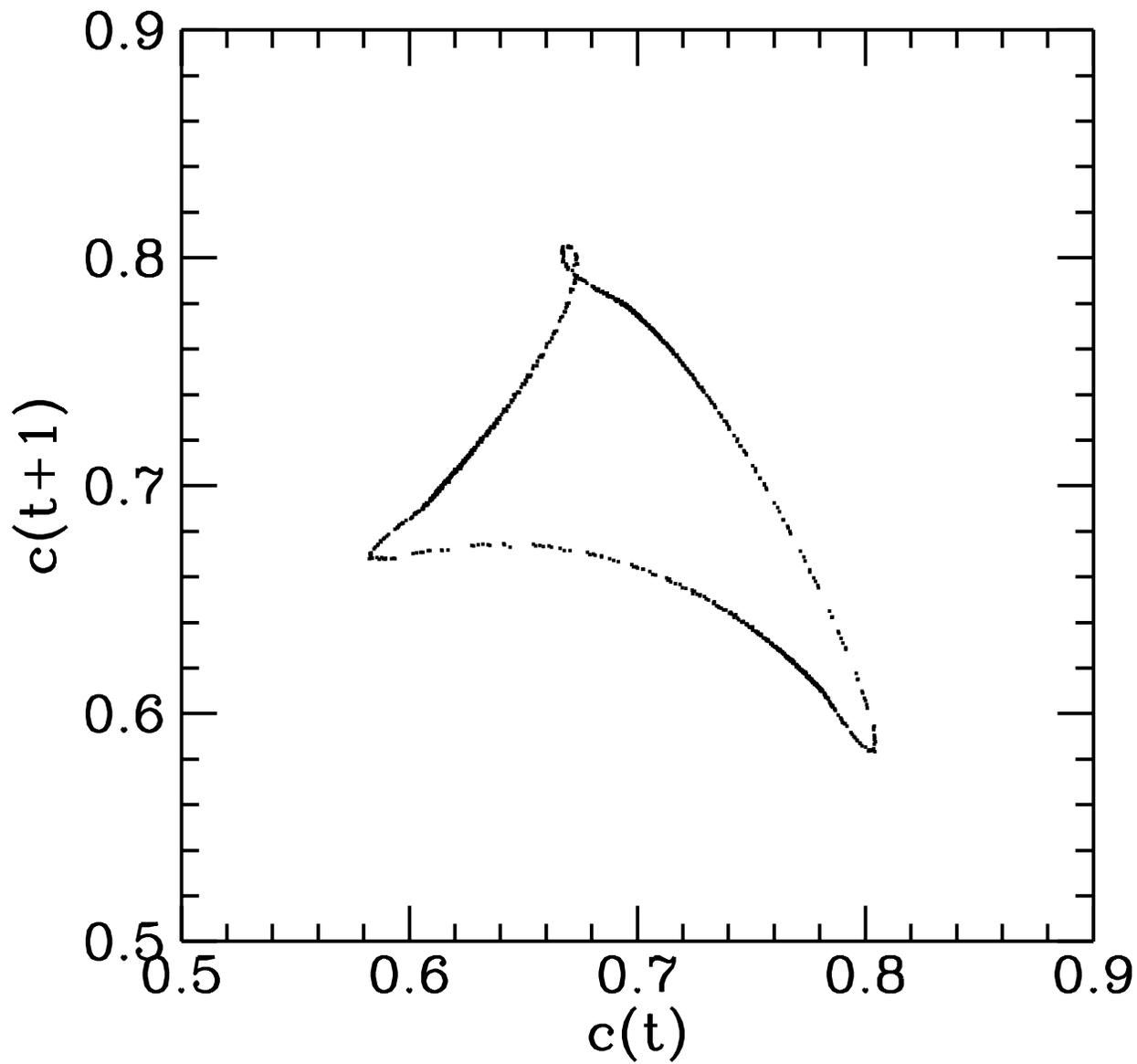
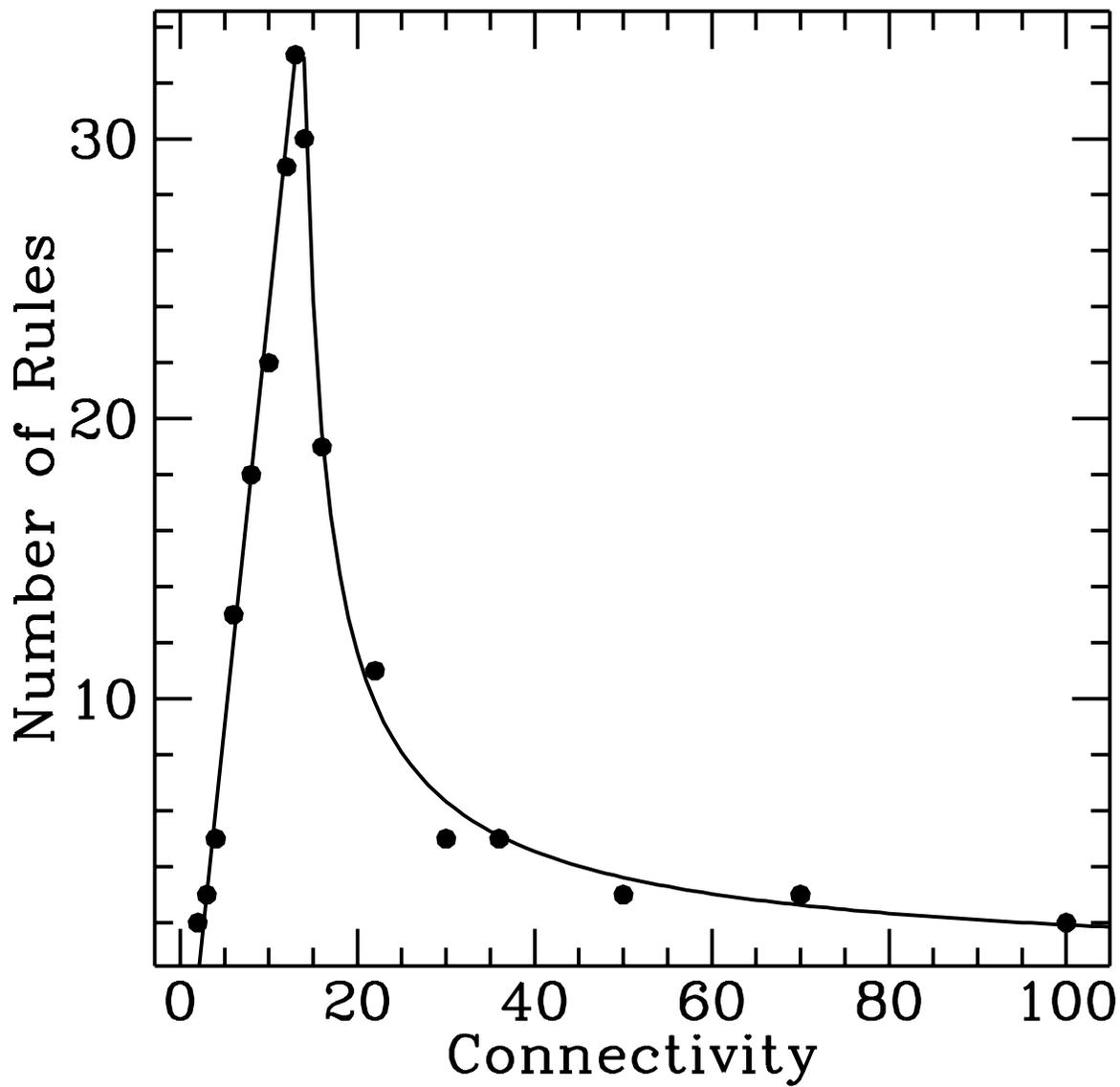
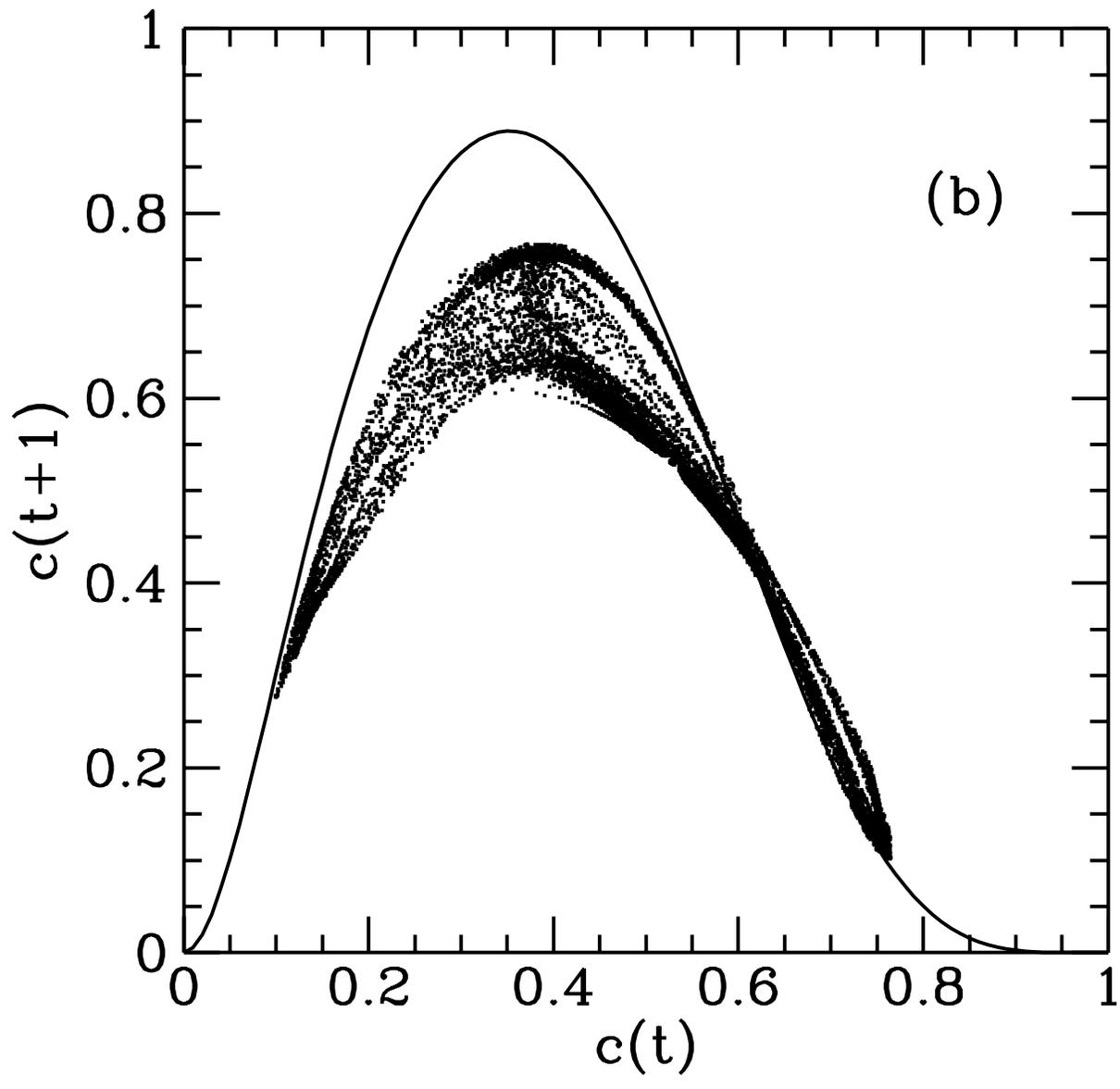
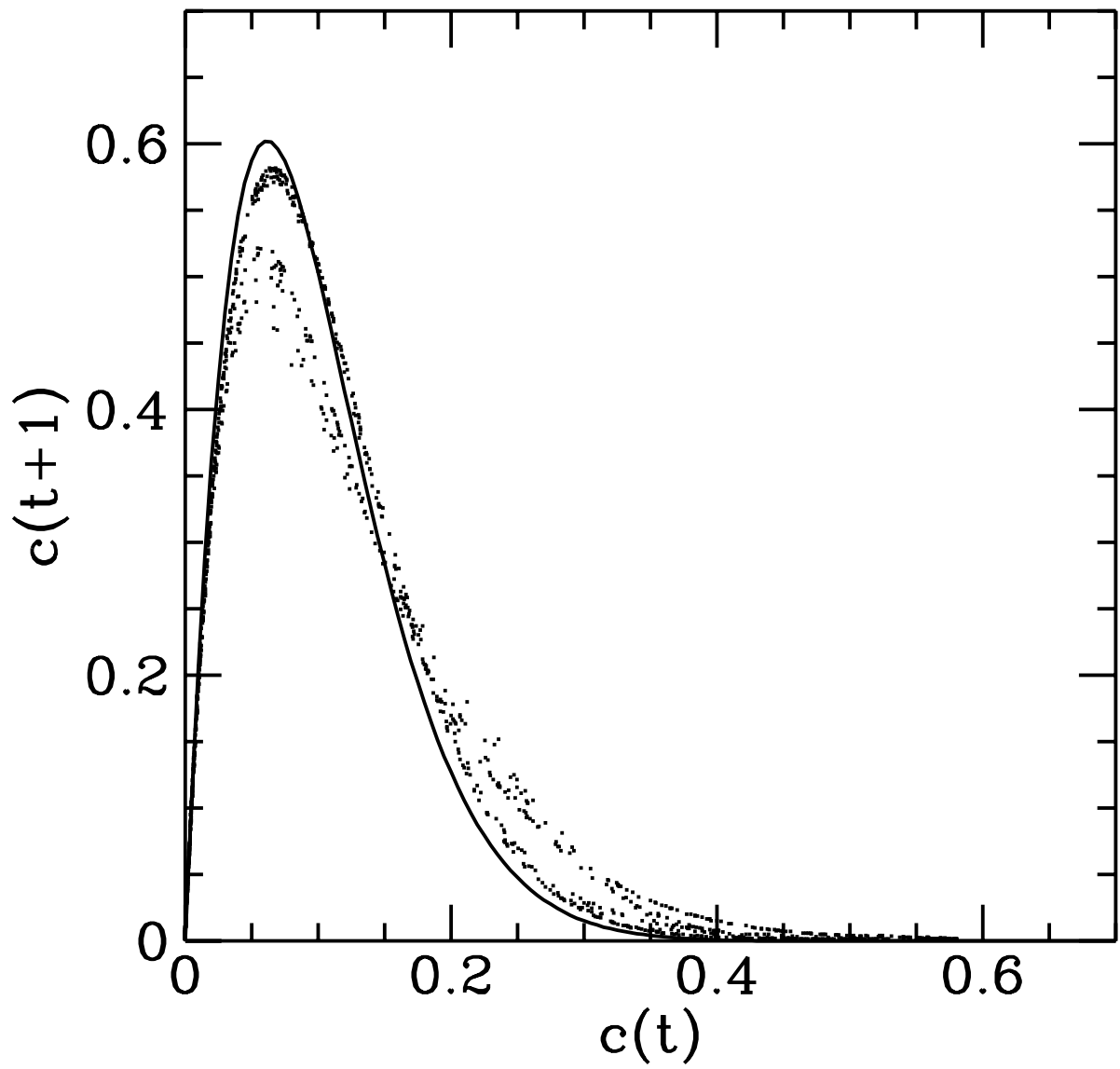
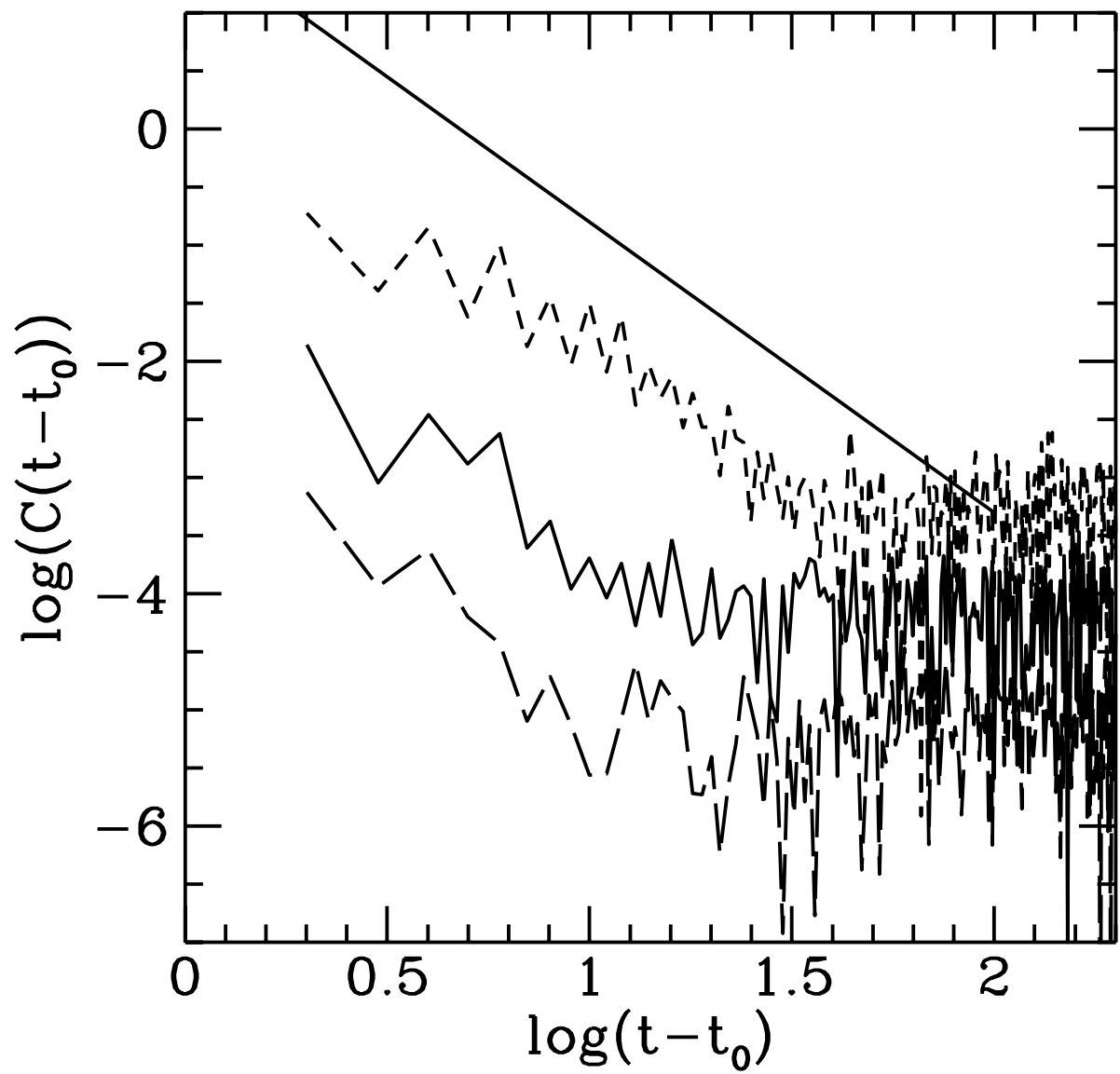
\end{document}